\documentclass{emulateapj}
\usepackage{graphicx}
\begin{document}

\def\etal{et al.\ \rm}

\title{Can giant planets form by direct gravitational instability?}

\author{R. R. Rafikov}
\affil{IAS, Einstein Dr., Princeton, NJ 08540}
\email{rrr@ias.edu}


\begin{abstract}
Gravitational instability has been invoked as a
possible mechanism of giant planet formation in protoplanetary disks.
Here we critically revise its viability by noting that for the direct 
production of giant planets it is not enough for protoplanetary 
disks to be gravitationally unstable. They must also
be able to cool efficiently (on a timescale comparable to the local
disk orbital period) to allow the formation of bound 
clumps by fragmentation. Combination of dynamical and thermal 
constraints puts very stringent lower limits on the 
surface density and temperature of disks capable of fragmenting  into
self-gravitating objects: for the gravitational instability to form giant planets 
at $10$ AU gas temperature at this location must
exceed $10^3$ K for a minimum disk mass of $0.7$ M$_\odot$ and 
minimum disk luminosity of $40$ L$_\odot$. Although these requirements 
relax in more distant parts of the disk, masses of bound 
objects formed as a result of instability even at $100$ AU are 
too large ($\sim 10$ M$_J$) to explain characteristics of
known extrasolar giant planets. Such protoplanetary disks 
(and planets formed in them) 
have very unusual observational properties and this 
severely constrains the possibility of giant planet formation by 
direct gravitational instability. 
\end{abstract}

\keywords{planets and satellites: formation --- 
solar system: formation --- planetary systems: protoplanetary disks}


\section{Introduction.}
\label{sect:intro}


Recent discoveries of Jupiter-like planets around solar-type 
stars have rejuvenated the interest in the issue of the 
origin of giant planets. Core instability model 
(Perri \& Cameron 1974; Harris 1978;  
Mizuno 1980) in which Jupiter-like planets
acquire their massive gaseous atmospheres 
by unstable gas accretion onto the preexisting massive solid 
cores has been one of the most fruitful ideas in this field. 
For rather long time this avenue of planet formation 
did not seem compatible with the short observed lifetimes 
($10^6-10^7$ yr) of protoplanetary disks because of the long 
time needed for the core accumulation. However, recent work by
Rafikov (2003) and Goldreich \etal (2004) has found
core formation time to be actually quite short, essentially 
removing the timescale issue from the core instability scenario.
Nevertheless, gravitational instability (hereafter GI) in 
the protoplanetary disk (Cameron 1978; Boss 1998) 
has been put forward as an alternative mechanism of giant 
planet formation. In this model massive gaseous disk becomes 
gravitationally unstable and rapidly 
fragments into a number of self-gravitating bound structures,
which further collapse to become giant planets. A number of 
recent hydrodynamical simulations (Boss 1998; Mayer \etal 2002) 
employing isothermal equation of state (EOS) have confirmed 
this general picture and provided ample support for the 
idea of planet formation by GI. The goal of this study is 
to constrain this avenue of planet formation by putting special 
emphasis on the conditions necessary for the actual 
fragmentation of protoplanetary disk 
into self-gravitating objects.


\section{Dynamical and thermal constraints.}
\label{sect:constraints}


Gravitational instability in a Keplerian disk can operate only 
when the gas sound speed $c_s$, surface density $\Sigma$
and local angular frequency $\Omega$ satisfy the condition
\begin{eqnarray}
Q\equiv\frac{\Omega c_s}{\pi G\Sigma}<Q_0.
\label{eq:dyn_constr}
\end{eqnarray}
(Safronov 1960; Toomre 1964). Here $Q$ is the so-called 
``Toomre $Q$''; sound speed is defined as 
$c_s\equiv(kT/\mu)^{1/2}$, where $T$ is the disk midplane 
temperature, $\mu$ is the gas molecular weight, and $k$ is a 
Boltzmann constant. Analytical arguments and results 
of numerical simulations suggest that $Q_0\approx 1$. 
Equivalent way
of formulating the condition for operation of GI is to 
require
\begin{eqnarray}
\rho>\frac{\Omega^2}{\pi G Q_0}=1.9\times 10^{-7}\mbox{g cm}^{-3}
a_{AU}^{-3}Q_0^{-1},
\label{eq:rho_constr}
\end{eqnarray}
where $\rho$ is a midplane gas density and  
$a_{AU}\equiv a/(1~\mbox{AU})$ is the distance from the central 
star scaled by $1$ AU.

Dynamical constraint (\ref{eq:dyn_constr}) is a 
necessary condition for GI to set in. However, even
when (\ref{eq:dyn_constr}) is fulfilled, giant 
planet formation becomes possible only if the disk 
can actually {\it fragment} into bound self-gravitating objects.
Recent studies (Gammie 2001; Rice \etal 2003) have demonstrated 
that fragmentation is possible only provided that the cooling time of 
the disk $t_{cool}$ satisfies
\begin{eqnarray}
\Omega t_{cool}<\xi,
\label{eq:thermal_constr}
\end{eqnarray}
where $\xi$ is a parameter of the order of unity; numerical 
simulations (Gammie 2001; Rice \etal 2003) suggest that 
$\xi\approx 3$. 

Cooling time of the disk can be estimated as the ratio
of its thermal energy to the escaping radiative flux:
\begin{eqnarray}
t_{cool}\approx\frac{\Sigma c_s^2}{\gamma-1}
\frac{f(\tau)}{\sigma T^4},~~~f(\tau)=\tau+\frac{1}{\tau},
\label{eq:t_cool}
\end{eqnarray}
where $\sigma$ is the Stephan-Boltzmann
constant, $T$ is the midplane disk temperature, 
$\tau\approx\kappa\Sigma/2$ is the disk optical depth
($\kappa$ is the opacity), and $\gamma$ is the adiabatic index of gas. 
Function $f(\tau)$ describes the efficiency of 
disk cooling and its specific form in (\ref{eq:t_cool}) 
corresponds to the case when cooling is radiative. 
When the disk is optically thick, $\tau\gg 1$, radiation has 
to leak out through the large optical 
depth of the disk material; this lowers the effective temperature
at the disk photosphere by a factor of $\tau^{1/4}$ compared 
to the midplane temperature $T$ and makes $t_{cool}$ very long
[first term in the definition of $f(\tau)$]. 
In the optically thin case, $\tau\ll 1$, according to the 
Kirchhoff's law (Rybicki \& Lightman 1979) disk emissivity 
is low and $t_{cool}$ is again very large [second term in 
the definition of $f(\tau)$]. Thus, disk cools most effectively
when $\tau\approx 1$. What is most important for our further 
discussion is that the value of $f(\tau)$ is above unity 
for any $\tau$ and any cooling mechanism because effective 
temperature of the disk cannot exceed its midplane 
temperature.

Dynamical constraint (\ref{eq:dyn_constr}) limits the value 
of the sound speed from above. At the same time, expressing
the temperature in (\ref{eq:t_cool}) through $c_s$ we find that
(\ref{eq:thermal_constr}) sets a lower limit on the gas sound 
speed. Combination of these conditions leads to the following
constraint on $c_s$ necessary for the giant planet formation 
by GI:
\begin{eqnarray}
\left[\Sigma\frac{f(\tau)}{\zeta}\frac{\Omega}{\sigma}
\left(\frac{k}{\mu}\right)^4\right]^{1/6}<c_s<\pi 
Q_0\frac{G\Sigma}{\Omega}, 
\label{eq:bound}
\end{eqnarray}
where $\zeta\equiv \xi(\gamma-1)\approx 1$.
Only when (\ref{eq:bound}) is fulfilled can the disk be 
gravitationally unstable and cooling be fast enough for 
fragmentation to allow the formation of bound gaseous clumps,
which later collapse to become giant planets.
Similar argument has been advanced by Levin (2003) in 
application to self-gravitating disks around AGNs.

At a specific location in the protoplanetary disk 
the condition (\ref{eq:bound}) can be satisfied only 
provided that the gas surface density obeys
\begin{eqnarray}
&& \Sigma>\Sigma_{min}\equiv \Sigma_{inf}[f(\tau)]^{1/5}
~~~\mbox{where}
\label{eq:sig_min}\\
&& \Sigma_{inf}\equiv\Omega^{7/5}(\pi G Q_0)^{-6/5}
\left[\frac{1}{\zeta\sigma}
\left(\frac{k}{\mu}\right)^4\right]^{1/5}\nonumber\\
&& \approx 6.6\times 10^5 \mbox{g cm}^{-2}a_{AU}^{-21/10}
\left(Q_0^{6}\tilde \mu^{4}\zeta\right)^{-1/5}
\left(\frac{M_\star}{M_\odot}\right)^{7/10}.
\label{eq:sigma_constr}
\end{eqnarray}
Here $\tilde \mu\equiv \mu/m_H$ is the mean molecular weight
relative to the atomic hydrogen mass  $m_H$ and $M_\star$ is 
the mass of the central 
star ($M_\odot$ is the Solar mass). For molecular gas of
solar composition $\tilde \mu\approx 2.3$ and $\Sigma_{min}
\approx 3.4\times 10^5$ g cm$^{-2}$ at $1$ AU [for 
$f(\tau)/(Q_0^{6}\tilde \mu^{4}\zeta)=1$], while if hydrogen 
is atomic $\tilde \mu\approx 1.2$ and $\Sigma_{min}
\approx 5.6\times 10^5$ g cm$^{-2}$. In both cases the disk surface 
density at 1 AU turns out to exceed that of the
minimum mass Solar nebula (Hayashi 1981) by more than 
$10^2$, implying that a very massive protoplanetary disk is 
necessary to sustain giant planet formation by GI. 

According to (\ref{eq:bound}), whenever 
(\ref{eq:sigma_constr}) is fulfilled, the sound speed in 
the disk is also bounded from below:
\begin{eqnarray}
&& c_s>c_{s,min}\equiv\Omega^{2/5}
\left[\frac{f(\tau)}{\zeta\pi Q_0 G\sigma}
\left(\frac{k}{\mu}\right)^4\right]^{1/5}\nonumber\\
&& \approx 6.9~ \mbox{km s}^{-1}a_{AU}^{-3/5}
\left[\frac{f(\tau)}{Q_0\tilde \mu^4\zeta}\right]^{1/5}
\left(\frac{M_\star}{M_\odot}\right)^{1/5}.
\label{eq:c_s_constr}
\end{eqnarray}
Because of that midplane temperature 
has to satisfy 
\begin{eqnarray}
&& T>T_{min}\equiv T_{inf}[f(\tau)]^{2/5}~~~\mbox{where}
\label{eq:T_min}\\
&& T_{inf}\equiv\Omega^{4/5}
\left(\zeta\pi Q_0 G\sigma
\right)^{-2/5}\left(\frac{k}{\mu}\right)^{3/5}\nonumber\\
&& \approx 5800~ \mbox{K}~a_{AU}^{-6/5}\tilde\mu^{-3/5}
\left(Q_0\zeta\right)^{-2/5}
\left(\frac{M_\star}{M_\odot}\right)^{2/5}.
\label{eq:T_constr}
\end{eqnarray}

In the form given by (\ref{eq:sig_min})-(\ref{eq:T_constr}) 
constraints on the disk surface density and temperature still 
depend on the behavior of opacity. However, bearing in mind that 
$f(\tau)>1$ for any $\tau$, one immediately sees that 
values of $\Sigma_{inf}$ and $T_{inf}$
\footnote{From  {\it infimum} 
--- the greatest lower bound of a set.} 
represent {\it absolute 
opacity-independent lower limits} on the disk surface density 
and temperature. Thus, we conclude that for the gravitationally
unstable disk to be able to fragment (which is necessary 
for giant planets formation) disk has to satisfy {\it at least}
$\Sigma>\Sigma_{inf}$ and $T>T_{inf}$. These lower limits
are extremely robust since they do not depend on a specific 
mechanism of energy losses from the disk --- by radiation or by 
convection. In practice, taking into account even the most basic 
properties of the energy transfer within the disk one can formulate 
much more stringent constraints, as we demonstrate below.


\section{Application to protoplanetary disks.}
\label{sect:application}


Protoplanetary disk luminosity $L_d$ and mass $M_d$ are 
the characteristics which can be 
directly observed (or constrained using observations).
According to (\ref{eq:sig_min}), mass of a protoplanetary 
disk which is capable of planet formation in the range of 
semimajor axes from $a_{in}$ to $a_{out}$ has to satisfy
\begin{eqnarray}
M_d>4.7M_\odot \left(a_{in,AU}^{-1/10}-a_{out,AU}^{-1/10}\right)
\tilde\mu^{-4/5}
\left(\frac{M_\star}{M_\odot}\right)^{7/10}.
\label{eq:M_d}
\end{eqnarray}
This and all subsequent numerical estimates assume 
$Q_0=1$ and $\zeta=1$.
Note a rather weak dependence of $M_d$ on the disk dimensions.
For the inner disk cutoff at $a_{in}=0.1$ AU and outer disk
cutoff at $a_{out}=100$ AU one finds that $M_d\gtrsim 3$ 
M$_\odot$ (for $\tilde\mu=1$). 
This estimate assumes maximum  
cooling efficiency throughout the whole disk, i.e $f(\tau)=1$,
but even such lowest possible limit on $M_d$ well exceeds the 
mass of the central star.

This constraint on the disk mass can be somewhat relaxed 
if disk is producing planets only locally. Indeed, the 
typical scale length of the fastest-growing perturbation in the  
marginally gravitationally unstable disk ($Q\approx 1$) 
is $\lambda\approx 2\pi h$, where
$h\equiv c_s/\Omega$ is the vertical disk scaleheight. 
Thus, to form planets by GI it is, in principle, 
sufficient that only an annulus of the disk with the radial 
width of order $\lambda$ has surface density and temperature 
exceeding $\Sigma_{min}$ and $T_{min}$.
Using (\ref{eq:c_s_constr}) we can estimate 
\begin{eqnarray}
&& \frac{h}{a}>\frac{1}{\Omega^{3/5}a}
\left[\frac{f(\tau)}{\zeta\pi Q_0 G\sigma}
\left(\frac{k}{\mu}\right)^4\right]^{1/5}\nonumber\\
&& \approx 0.23~a_{AU}^{-1/10}
\left[\frac{f(\tau)}{\tilde \mu^{4}}\right]^{1/5}
\left(\frac{M_\star}{M_\odot}\right)^{-3/10}.
\label{eq:h}
\end{eqnarray}
Because of the high midplane temperature 
(necessary to ensure efficient cooling) disk is not
very thin geometrically; as a result, GI is likely
to have global character. Mass of such a ``minimum 
planet-forming annulus'' centered on $a$ and having 
width $\lambda$ is  constrained by
\begin{eqnarray}
M_{a} \gtrsim 0.8 M_\odot a_{AU}^{-1/5}
\left[f(\tau)\right]^{2/5}
\tilde \mu^{-8/5}
\left(\frac{M_\star}{M_\odot}\right)^{2/5}.
\label{eq:M_a}
\end{eqnarray}
Clearly, this lower limit on $M_a$ depends only very weakly 
on the semimajor axis of annulus. It also amounts to a sizable 
fraction of 
$M_\star$ [even for molecular gas with $\tilde \mu =2.3$ 
and $f(\tau)=1$ one finds $M_a\gtrsim 0.2$ M$_\odot$ at $1$ AU].
At the same time, typical masses of protoplanetary disks 
inferred from mm and infrared 
observations vary between $10^{-3}$ 
M$_\odot$ and $0.1$ M$_\odot$ (Kitamura \etal 2002), but these masses of gas 
are typically extended over $\sim 100$ AU in disk radius (and 
not within just a narrow annulus). Besides, in the real 
nebula such annulus can easily be either very optically thick or
very optically thin, which translates into large value of $f(\tau)$, 
additionally increasing the lower limit on $M_a$ (see below).  

Energy flux emitted from both sides by a unit surface area 
of the disk is given by 
\begin{eqnarray}
\frac{dL_d}{dS}=2\frac{\sigma T^4}{f(\tau)}>
2\Omega^{16/5}\left[\frac{f(\tau)}{\sigma}\right]^{3/5}
\left[\frac{(k/\mu)^{3/2}}{\pi G}\right]^{8/5}
\label{eq:emissivity}
\end{eqnarray}
where we made use of (\ref{eq:T_constr}). Since  
$dL_d/dS\propto a^{-24/5}$ most of the 
energy is radiated by the innermost part of  
the gravitationally unstable disk. Total 
luminosity of a disk with an inner cutoff at $a$ is
\begin{eqnarray}
&& L_d=2\times 2\pi\int\limits_{a}\frac{dL_d}{dS}ada\approx 
\frac{10\pi}{7}a^2\frac{dL_d}{dS}(a)>\nonumber\\
&& 1.6\times 10^4 L_\odot~a_{AU}^{-14/5}\left[\frac{f(\tau)}{\tilde\mu^4}\right]^{3/5}
\left(\frac{M_\star}{M_\odot}
\right)^{8/5}.
\label{eq:L_d}
\end{eqnarray}
This estimate of luminosity holds even if only an 
annulus of gas is considered instead of the full disk, which 
is a direct consequence of the very steep  dependence of 
$dL_d/dS$ on the distance from the central star. 

We now consider what these limits imply  for the disk properties at 
different locations in the protoplanetary nebula.


\subsection{Limits on the disk properties at $1$ AU.}
\label{subsect:1_AU}


First we consider the possibility of giant planet formation
by GI in the region of terrestrial planets, at $a=1$ AU. 
Constraints obtained in previous sections imply that planetary 
genesis at this location requires rather extreme disk properties: 
temperature 
has to exceed at least $T_{inf}=5.2\times 10^3$ K (this estimate is 
based using $\tilde \mu=1.2$ in [\ref{eq:T_constr}] 
since gas cannot be molecular at such temperature), surface 
density must be above $\Sigma_{inf}=5.7\times 10^5$ g cm$^{-2}$, and 
disk luminosity has to exceed $10^4$ L$_\odot$
(all assuming $\zeta=1$ and $Q_0=1$). 
Depending on whether an extended disk or just a minimum size 
annulus is considered, minimum mass varies from $M_d\approx 1.5$ M$_\odot$
to $M_a\approx 0.6$ M$_\odot$. Apparently, even these least
radical requirements are in complete disagreement with the observed
properties of protoplanetary disks. 

The argument can be significantly sharpened by noticing that the 
disk with such high surface density has to be optically very 
thick, meaning that $\Sigma_{min}$ and $T_{min}$ provide much 
more stringent limits on the disk properties than $\Sigma_{inf}$ 
and $T_{inf}$. Using opacity dependence $\kappa(\rho,T)$ 
on the temperature $T$ and gas density $\rho$ 
(given by [\ref{eq:rho_constr}]) 
from Bell \& Lin (1994) we can substitute 
$\tau=\kappa(\rho,T)\Sigma_{min}/2$ into (\ref{eq:sig_min}) and 
(\ref{eq:T_min}) and solve the resulting system for $\Sigma_{min}$
and $T_{min}$. Performing this procedure  one finds that at $1$ AU 
disk has to be extremely hot so that opacity is due to 
electron scattering. Disk temperature has to exceed
$10^6$ K, but this is impossible, because such disk
would not be bound to the central star and its radiation
pressure would far dominate over the gas pressure.
This firmly rules out the possibility of giant planet 
formation by GI within several AU from the central star.


\subsection{Limits on the disk properties at $10$ AU.}
\label{subsect:10_AU}


At $10$ AU, in the region of giant planets, temperature will
probably be low enough for the gas to be molecular. Using
$\tilde\mu=2.3$ we find $\Sigma_{inf}=2.7\times 10^3$ g cm$^{-3}$
and $T_{inf}=220$ K. Disk luminosity has 
to exceed only $3.4$ L$_\odot$, and the full disk and minimum mass 
annulus have to contain at least $M_d=0.4$ M$_\odot$ 
(for $a_{out}=100$ AU) and $M_a=0.13$ M$_\odot$ correspondingly.
These limits are more reasonable than at $1$ AU, 
although mass is still high and disk is too hot compared 
to the observed systems, which typically have $M_d<0.1$ M$_\odot$ 
and $T\lesssim 10^2$ K at $10$ AU (e.g. Kitamura \etal 2002).

However, using again opacities from Bell \& Lin (1994) and 
$\rho$ as given by (\ref{eq:rho_constr}) we find that real disks
(i.e. cooling not at the maximum efficiency)
should be considerably more extreme. There are several possible 
solutions for $T_{min}$ and $\Sigma_{min}$ at $10$ AU
corresponding to different opacity regimes. The least extreme one 
is a "cold" solution with $\tau\approx 60$, 
$T_{min}\approx 1100$ K, $\Sigma_{min}\approx 6\times 10^3$ 
g cm$^{-2}$, $L_d\gtrsim 40$ L$_\odot$, $M_a\gtrsim 0.7$ M$_\odot$
which corresponds to molecular gas at the temperature of
grain evaporation (other solutions have 
$T_{min}\gtrsim 7\times 10^3$ K). This again yields a minimum 
disk which is too massive and hot to satisfy current 
observational constraints. Thus, it is extremely unlikely 
that GI can allow disk fragmentation and
subsequent giant planet formation even at $10$ AU.


\subsection{Limits on the disk properties at $100$ AU.}
\label{subsect:100_AU}


We also look at the possibility of giant planet formation 
by GI in the distant regions of protoplanetary nebula, 
at $a=100$ AU. For molecular disk at this location we 
find $\Sigma_{inf}=20$
g cm$^{-3}$, $T_{inf}=14$ K, $L_d\gtrsim 5\times 10^{-3}$
L$_\odot$, and $M_a\gtrsim 0.08$ M$_\odot$. 
These properties change only a little when inefficiency of 
disk cooling is properly accounted for: using 
$\kappa\approx 0.1(T/10~\mbox{K})$ cm$^2$ g$^{-1}$ in
agreement with observations of protoplanetary nebulae
(Beckwith \etal 1990; Kitamura \etal 2002) we find
that to be able to form giant planets by GI
disk has possess at least $\tau\approx 2$ (marginally 
optically thick), 
$T_{min}\approx 20$ K, $\Sigma_{min}\approx 25$ g cm$^{-2}$, 
$L_d\gtrsim 10^{-2}$ L$_\odot$, and 
$M_a\gtrsim 0.1$ M$_\odot$ at $100$ AU. 
Although $M_a$ is 
very near the upper end of the observed distribution of 
protoplanetary disk masses, these parameters seem
to be acceptable from the observational point of view. 
There are however additional reasons to doubt the possibility 
of planet formation
by GI even at $100$ AU, which we discuss next. 


\subsection{Fragment masses.}
\label{subsect:frag_mass}


Another important observational constraint on the planet
formation by GI comes from comparing the observed masses 
of extrasolar giant planets with the typical masses of fragments
into which disk breaks up when $t_{cool}<\xi\Omega^{-1}$. As we 
mentioned earlier, the lengthscale of the most unstable mode
is $\lambda\approx 2\pi h$, which results in the minimum 
fragment mass of 
\begin{eqnarray}
M_f\approx \Sigma_{min}\lambda^2\approx 0.15~\mbox{M}_\odot
a_{AU}^{-3/10}\left[\frac{f(\tau)}{\tilde\mu^{4}}\right]^{3/5}
\left(\frac{M_\star}{M_\odot}
\right)^{1/10}.
\label{eq:M_f}
\end{eqnarray}
At $100$ AU molecular disk meeting the requirements 
outlined in \S \ref{subsect:100_AU} would fragment into
self-gravitating clumps with the mass of roughly 
$5$ M$_J[f(\tau)]^{3/5}$, where $M_J$ is the Jupiter's mass;
at smaller semimajor axes $M_f$ would be higher.
Even for $f(\tau)=1$ this mass is larger than masses of 
most extrasolar giant planets detected to date (Marcy \etal 2003). 
More realistic cooling efficiency
corresponding to the optical depth of
$\tau\approx 2$ at $100$ AU leads to  
$M_f\approx 9$ M$_J$, landing
minimum fragment mass not too far from 
the brown dwarf regime. 

It is possible that such clumps would be able to further 
fragment into smaller objects but we view this outcome 
as rather unlikely. Indeed, as the fragment contracts its
optical depth increases and cooling time in this nonlinear 
regime becomes larger than the dynamical time of collapsing
fragment, which disfavors subsequent fragmentation 
(Goodman \& Tan 2004). It remains
to be seen how the centrifugal support of the collapsing
clump can change this conclusion.


\section{Discussion.}
\label{sect:disc}


Novel analytical constraints 
presented in \S \ref{sect:constraints} and \ref{sect:application}, 
when confronted with observations of protoplanetary disks, 
severely undermine the possibility of giant planet 
formation by GI. In particular, we have demonstrated that 
disks capable
of producing giant planets by GI at a distance of several AU
from the central star cannot exist simply on dynamical 
grounds --- to cool efficiently they must be too hot to
be bound to the central object. This essentially rules out 
a possibility of in situ formation of close-in extrasolar 
giant planets ("hot Jupiters") by GI. Rafikov (2004) 
have previously presented arguments against in situ 
formation of ``hot Jupiters'' via the core 
instability. Then the most natural way to explain the 
existence of close-in giant planets is 
to accept that they have formed elsewhere under more
favorable conditions and then migrated to their current locations.

Planet formation by GI is also extremely unlikely within several
tens of AU, as results of \S \ref{subsect:10_AU} demonstrate:
disks with required properties must be so hot ($T\gtrsim 10^3$ K), 
luminous (several tens of $L_\odot$), and massive ($\sim M_\odot$) that
they would clearly stand out in a sample of observed protoplanetary
nebulae. Beyond about $100$ AU minimum disk properties allowing planets
to form by GI become roughly acceptable from the observational 
point of view, although disk masses still reside at the very upper 
end of the observed distribution of protoplanetary nebulae masses. 
However, it is still very unlikely that the extrasolar giant planets that
we see now could have been produced by GI even at several hundred 
AU from the central star. The problem is not only in a potential 
difficulty of migrating such planets from beyond $100$ AU all 
the way in to several AU, but also in producing planets with the 
right mass. As our estimate (\ref{eq:M_f}) of the minimum mass 
of unstable disk fragments demonstrates, bound objects produced 
by GI even at $100$ AU are too massive ($\sim 10$ M$_J$) to explain
the observed mass distribution of extrasolar planets.  

Our study emphasizes the importance of the proper treatment of disk
thermodynamics (especially its cooling) for studying the 
possibility of Jupiter-like planet formation. By now 
virtually all simulations which were able to demonstrate
disk fragmentation and collapse of resulting dense  
objects in gravitationally unstable disks used isothermal 
EOS (e.g. Mayer \etal 2002). However, use of this EOS is 
equivalent to setting the disk cooling time to zero which 
artificially relaxes the requirements for the planet 
production process and is misleading.  Not surprisingly, more
realistic calculations following thermodynamics in greater 
detail typically do not exhibit fragmentation of gravitationally 
unstable disks which are not capable of cooling efficiently 
(Pickett \etal 1998; Gammie 2001; 
Rice \etal 2003). Thus, simulations employing 
isothermal EOS should not be trusted too much when planet 
formation in real 
protoplanetary disks is concerned. 

Future infrared and mm observation will show whether 
protoplanetary disks with extreme properties satisfying the  
constraints necessary for giant planet formation by GI 
really exist.

\acknowledgements 

I am grateful to Yuri Levin, who has been working on similar 
subjects, and Lynne Hillenbrand for illuminating discussions.
Author is a Frank and Peggy Taplin Member at the IAS;
he is also supported by the 
W. M. Keck Foundation and NSF grant PHY-007092.


\end{document}